\documentclass[showpacs,showkeywords,aps,graphicx,twocolumn]{revtex4}
\usepackage{graphicx}
\usepackage{array}
\usepackage{subfigure}
\begin{document}

\title{Measurement-device-independent quantum key distribution of multiple degrees of
freedom of a single photon}

\author{ Yu-Fei Yan$^{1}$, Lan Zhou$^{2}$, Wei Zhong$^{1,3}$,  Yu-Bo Sheng$^{1,3}$ \footnote{Corresponding author: shengyb@njupt.edu.cn}}

\address{$^1$ Institute of Quantum Information and Technology, Nanjing University of Posts and Telecommunications, Nanjing, 210003, China\\
 $^2$ Schoool of Science, Nanjing University of Posts and Telecommunications, Nanjing,
210003, China\\
 $^3$ Key Lab of Broadband Wireless Communication and Sensor Network Technology, Nanjing University of Posts and Telecommunications, Ministry of
 Education, Nanjing, 210003, China\\
Corresponding authors E-mail: shengyb@njupt.edu.cn\\}

\begin{abstract}
Measurement-device-independent quantum key distribution (MDI-QKD) provides us a powerful approach to resist all attacks at detection side. Besides the unconditional security, people also seek for high key generation rate, but MDI-QKD has relatively low key generation rate. In this paper, we provide an efficient approach to increase the key generation rate of MDI-QKD by adopting multiple degrees of freedom (DOFs) of single photons to generate keys. Compared with other high-dimension MDI-QKD protocols encoding in one DOF, our protocol is more flexible, for our protocol generating keys in independent subsystems and the detection failure or error in a DOF not affecting the information encoding in other DOFs. Based on above features, our MDI-QKD protocol may have potential application in future quantum communication field.
\end{abstract}

\keywords{Measurement-device-independent quantum key distribution, polarization, longitudinal-momentum, key generation rate}

\pacs{03.67.Pp, 03.67.Hk, 03.65.Ud}

\maketitle

\section{Introduction}
 Since the Bennett-Brassard-1984 (BB84) quantum key distribution (QKD) protocol \cite{yinyong2} was proposed,
quantum secure communication based on QKD has experienced a long-term development, which has become an interesting research area in both theory and experiment  \cite{ekert1991quantum,lo1999,cerf2002security,stucki2002quantum,
 grosshans2003quantum,lo2005,schmitt2007experimental,Koashi2009,wang2018,fp,wang2019,xu2020quantum,qe,qe1,qe2}. Throughout the whole development process of QKD, there are always struggle between attack and anti-attack strategies, which focus on the discrepancy between perfect theory and practice \cite{yinyong4}. Although QKD is absolutely secure in theory, in reality, due to the imperfect measurement devices and quantum signal source, QKD system cannot guarantee the unconditional security of keys in practical condition.
 For example, practical photon detectors are vulnerable to various types of attacks such as the time-shift attacks \cite{timeshifitattcak5,timeshifitattcak6,timeshifitattcak7}, detection blinding attacks \cite{blindattacks2}, beam-splitter attack \cite{splitter1,splitter2} and fate-stack attacks \cite{fate-stackattacks9,fate-stackattacks10}. In addition, eavesdroppers can also take use of imperfect quantum signal sources to perform attack \cite{signalsourcesattack7,signalsourcesattack8}, such as the photon-number-splitting (PNS) attack \cite{pnsattcak9}. The problem from PNS attack can be solved by the decoy state method \cite{decoystates10,decoystates11,decoystates12,decoystates13}.

In 2007, Ac\'{i}n group first put forward the device-independent QKD (DI-QKD) protocols \cite{DI-QKD1}. DI-QKD does not require detailed knowledge of how devices work and ensures the security based on the violation of Bell inequality. DI-QKD can defend all possible attacks from imperfect experimental devices. However, DI-QKD requires quite high detection efficiency, which limits its secure communication distance to be less than 5 $km$ \cite{DI-QKD2,DI-QKD3,DI-QKD4}.
In 2012, the group of Lo proposed the concept measurement-device-independent QKD (MDI-QKD), which is immune to all possible attacks from measurement devices \cite{MDI-QKD}. In MDI-QKD, Alice and Bob prepare single photons randomly in one of two bases and send the encoding photons to a third party (Charlie). Charlie performs a Bell-state analysis (BSA) on them and publish his result. Here, Charlie needs not to be trusted and can even be completely controlled by the eavesdropper. Combined with the decoy state method, the MDI-QKD can resist all the possible attacks from imperfect photon source and measurement devices. During the past few years,
 MDI-QKD has well developed \cite{MDI-QKD1,MDI-QKD2,MDI-QKD3,MDI-QKD7,MDI-QKD4,MDI-QKD6,MDI-QKD8,MDI-QKD9,MDI-QKD10}. For example, in 2016, Yin \emph{et al.} reported results for MDI-QKD over 404 km of ultralow-loss optical fiber and 311 km of standard optical fiber \cite{MDI-QKD7}. Besides security, practical application requires high key generation rate. However, comparing with other QKD forms, MDI-QKD has relatively low key generation rate. Increasing single photon's capacity is important for increasing MDI-QKD's key generation rate. In general, there are two methods to increase single photon's capacity. The first one is the high-dimension (HD) encoding in one degree of freedom (DOF) \cite{highdimension1,highdimension2,highdimension3,highdimension4,highdimension5,highdimension6,chao,highdimension8,
 highdimension9,highdimension10,highdimension11,highdimension12,highdimension13}. In 2016, the group of Chau proposed the first qudit-based MDI-QKD protocol using
linear optics \cite{chao}. In 2018, Dellantonio \emph{et al.} designed
two HD MDI-QKD protocols based on spatial-encoding and
temporal-encoding multi-mode qudits, respectively \cite{highdimension10}. The other one is the hyper-encoding in multiple DOFs \cite{multiQKD1,multiQKD2,multiQKD3}. For hyper-encoding, information is simultaneously encoded in multiple DOFs of a quantum
system. The information encoding
in different DOFs are independent with each other. The single photons encoded in multiple DOFs have also been used to realize high-capacity quantum communication, such as high-capacity quantum teleportation \cite{hypercommun2}, high-capacity dense coding \cite{hypercommun4}, and high-capacity quantum secure direct communication \cite{hypercommun6,hypercommun7,hypercommun8}.

   In this protocol, we propose a general method to increase MDI-QKD's key generation rate by adopting the qudits encoding in multiple DOFs to generate keys. We take the qudits encoding in polarization and two longitudinal momentum DOFs for example, which have been experimentally generated \cite{generation}. In this protocol, both Alice and Bob encode their photons in three DOFs independently and send the encoded photons to Charlie for the hyperentanglement Bell state analysis (HBSA). In our MDI-QKD protocol, the encoding information in all DOFs can be used to generate the secure key, so that it can effectively increase MDI-QKD's key generation rate. Each DOF can be manipulated independently and they do not influence each other.
 We will prove that our MDI-QKD protocol is unconditionally secure and calculate its key generation rate. Our MDI-QKD protocol  has big application potential in future quantum secure communication field.

The structure of the paper is organized as follows. In section II, we explain MDI-QKD protocol based on the qudits encoding in polarization and two longitudinal momentum DOFs.
 In section III, we make the security analysis of our MDI-QKD protocol and evaluate its key generation rate. In section IV, we give a discussion. In Section V, we make a conclusion.

\section{The MDI-QKD protocol adopting three DOFs of single photons}
Before we begin our protocol, we first introduce our encoding rules in three DOFs. In each DOF, Alice and Bob use two sets of bases, say, rectilinear base and diagonal base. In the polarization DOF, the rectilinear base is $\{|H\rangle,|V\rangle\}$, where $|H\rangle$ and $|V\rangle$ denote horizontal and vertical polarizations, respectively. The diagonal base is $\{|+\rangle _{p}=\frac{1}{\sqrt{2}}(|H\rangle+|V\rangle)$,
$|-\rangle _{p}=\frac{1}{\sqrt{2}}(|H\rangle-|V\rangle)\}$. In the first longitudinal-momentum DOF, the rectilinear base is \{$|L\rangle$,$|R\rangle$\}, where $|L\rangle$ and $|R\rangle$ represent the left and right spatial modes. The diagonal base includes  \{$|+\rangle _{f}=\frac{1}{\sqrt{2}}(|L\rangle+|R\rangle)$,
$|-\rangle _{f}=\frac{1}{\sqrt{2}}(|L\rangle-|R\rangle)$\}. In the second longitudinal-momentum DOF, the rectilinear base is $\{|I\rangle$,$|E\rangle\}$ and the diagonal base is $\{|+\rangle _{s}=\frac{1}{\sqrt{2}}(|I\rangle+|E\rangle)$, $|-\rangle _{s}=\frac{1}{\sqrt{2}}(|I\rangle-|E\rangle)\}$, where $|I\rangle$ and $|E\rangle$ represent the internal and external spatial modes, respectively.  In polarization DOF, $|H\rangle$ and $|+\rangle _{p}$ represent the binary value 0 while $|V\rangle$ and $|-\rangle _{p}$ represent the binary value 1. In the first (second) longitudinal-momentum DOF, $|L(I)\rangle$ and $|+\rangle _{f(s)}$ represent 0 while $|R(E)\rangle$ and $|-\rangle _{f(s)}$ represent 1.

The basic principle of our MDI-QKD protocol is shown in Fig. 1, whose procedure can be described as follows.
\begin{figure*}
\centering
\subfigure[]{
\begin{minipage}[t]{0.4\linewidth}
\centering
\includegraphics[scale=0.3]{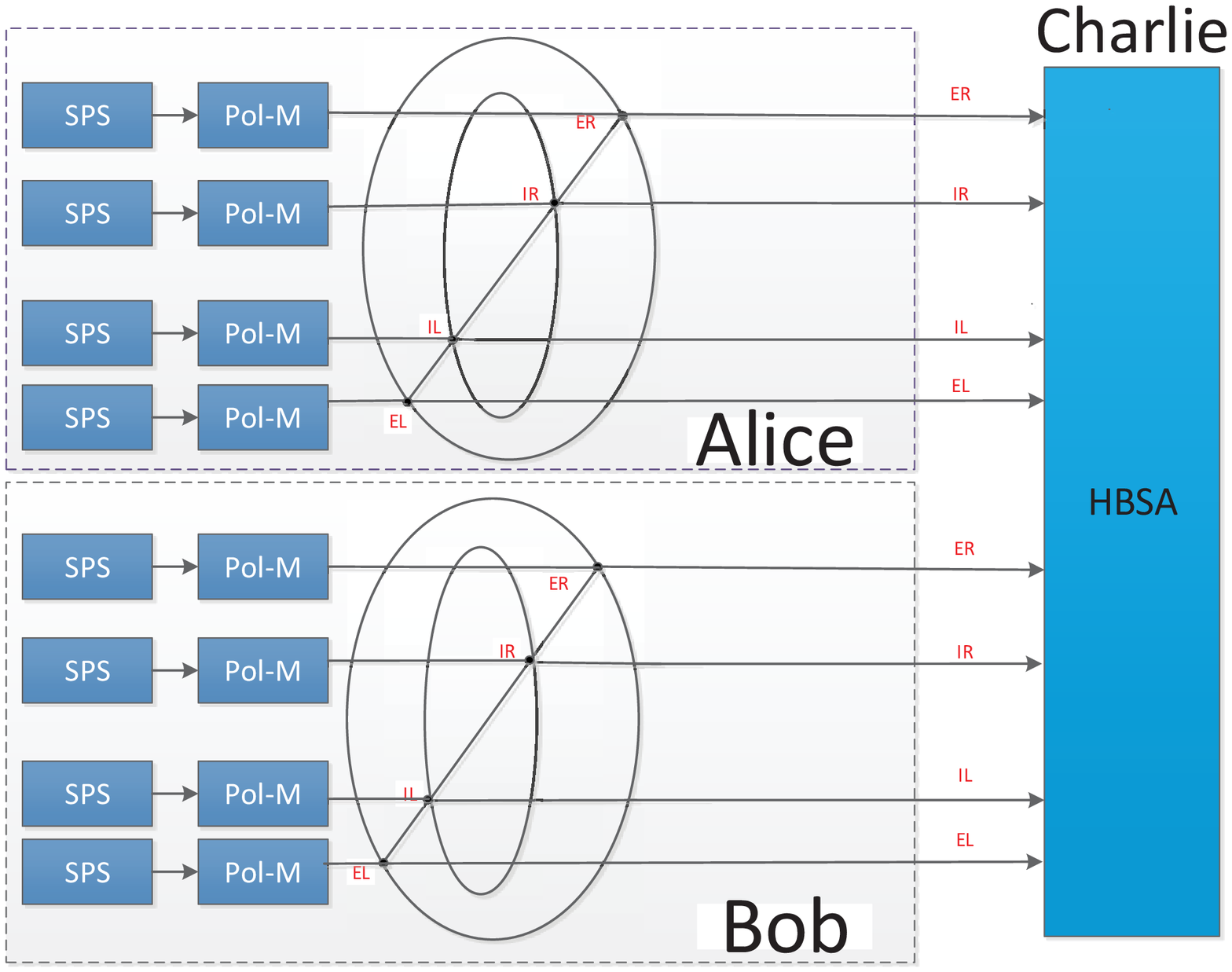}
\end{minipage}
}
\subfigure[]{
\begin{minipage}[t]{0.4\linewidth}
\centering
\includegraphics[scale=0.3]{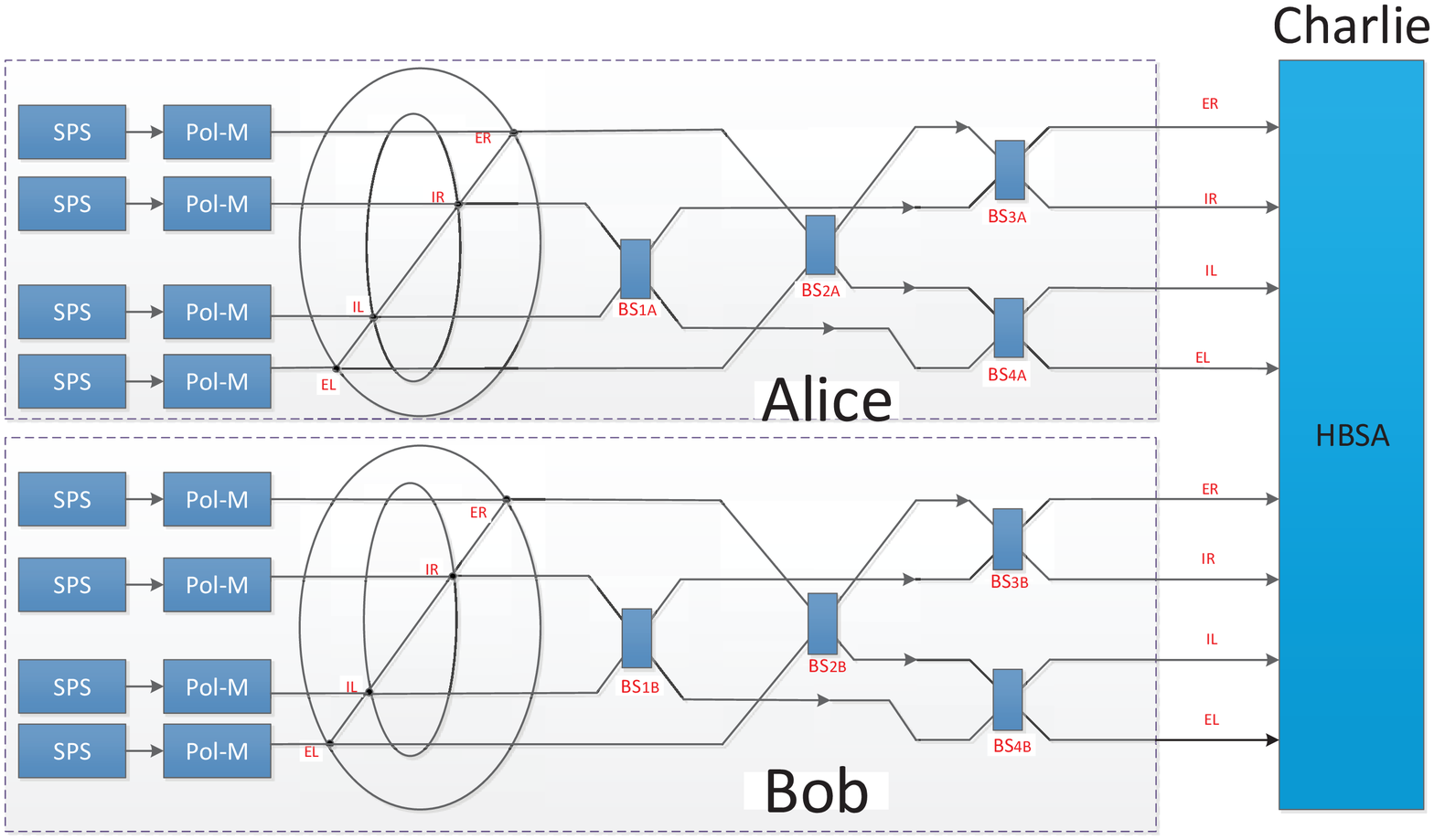}
\end{minipage}
}
\centering
\caption{Schematic diagram of the MDI-QKD protocol adopting three DOFs of single photons.  Alice and Bob separately prepare single photons from single photon sources (SPSs). Pol-M is a polarization modulator which is used to implement encoding in polarization DOF. (a) and (b) represent the encoding in rectilinear and diagonal bases in two longitudinal momentum DOFs, respectively. The BSs are used to perform a Hadamard operation in two longitudinal-momentum DOFs. Here, $|L\rangle$ ($|R\rangle$) and $|I\rangle$ ($|E\rangle$) represent the left (right) and internal (external) momentums. After the qudits generation, the qudits from Alice and Bob are sent to Charlie to make the complete hyperentangled-Bell-state analysis (HBSA), which can completely distinguish the four Bell states in each DOF.}
\end{figure*}

Step1: Alice and Bob separately prepare single photons from single photon sources (SPSs). First, they make the single photons pass through polarization modulators (Pol-Ms) to realize the encoding in polarization DOF. In Fig. 1(a) and (b), we show the generation principle of qudits in rectilinear and diagonal bases in both longitudinal momentum DOFs, respectively. BS means the 50:50 beam splitter and acts as the Hadamard gate in two longitudinal-momentum DOFs. It can convert $|L\rangle$ into $|+\rangle _{f}$, $|R\rangle$ into $|-\rangle _{f}$, $|I\rangle$ into $|+\rangle _{s}$, and $|E\rangle$ into {$|-\rangle _{s}$}. Here, we give a simple example for the generation of qudits in the diagonal base of both two longitudinal-momentum DOFs. As shown in Fig. 1(b), if Alice (Bob) wants to generate $|+\rangle_{f}\otimes|+\rangle_{s}$, she (he) first prepares a single photon in the spatial momentum $|L\rangle|E\rangle$. Then, after passing the photon successively through $BS_{2A(B)}$, $BS_{3A(B)}$, and $BS_{4A(B)}$, the photon state $|L\rangle|E\rangle$ will finally evolve to $|+\rangle _{f}\otimes|+\rangle _{s}$. In this way, Alice and Bob can prepare single-photon qudits encoding in three DOFs. For instance, if Alice's encoding key is 000, there are totally eight possible states that Alice can prepare, say, {$|H\rangle\otimes|L\rangle\otimes|I\rangle$}, {$|H\rangle\otimes|L\rangle\otimes|+\rangle _{s}$}, {$|H\rangle\otimes|+\rangle _{f}\otimes|I\rangle$}, {$|H\rangle\otimes|+\rangle _{f}\otimes|+\rangle _{s}$}, {$|+\rangle _{p}\otimes|L\rangle\otimes|I\rangle$}, {$|+\rangle _{p}\otimes|L\rangle\otimes|+\rangle _{s}$}, {$|+\rangle _{p}\otimes|+\rangle _{f}\otimes|I\rangle$}, and
{$|+\rangle _{p}\otimes|+\rangle _{f}\otimes|+\rangle _{s}$}. After encoding, Alice and Bob send their qudits to a third party Charlie through quantum channels, respectively.

Step2: After Charlie receiving the photon qudits sent by Alice and Bob, he performs a hyperentangled-Bell-state analysis (HBSA) \cite{HBSA} on the photon pairs, which can completely distinguish the four Bell states in each DOF. The Bell states in three DOFs can be written as
\begin{eqnarray}
|\Phi^{\pm}\rangle _{p}&=&\frac{1}{\sqrt{2}}(|H\rangle_{a}|H\rangle_{b}\pm|V\rangle_{a}|V\rangle_{b}),\nonumber\\
 |\Psi^{\pm}\rangle _{p}&=&\frac{1}{\sqrt{2}}(|H\rangle_{a}|V\rangle_{b}\pm|V\rangle_{a}|H\rangle_{b}).\nonumber\\
|\Phi^{\pm}\rangle _{f}&=&\frac{1}{\sqrt{2}}(|L\rangle_{a}|L\rangle_{b}\pm|R\rangle_{a}|R\rangle_{b}),\nonumber\\
\qquad|\Psi^{\pm}\rangle _{f}&=&\frac{1}{\sqrt{2}}(|L\rangle_{a}|R\rangle_{b}\pm|R\rangle_{a}|L\rangle_{b}),\nonumber\\
|\Phi^{\pm}\rangle _{s}&=&\frac{1}{\sqrt{2}}(|I\rangle_{a}|I\rangle_{b}\pm|E\rangle_{a}|E\rangle_{b}),\nonumber\\
\qquad|\Psi^{\pm}\rangle _{s}&=&\frac{1}{\sqrt{2}}(|I\rangle_{a}|E\rangle_{b}\pm|E\rangle_{a}|I\rangle_{b}),
\end{eqnarray}
where the subscript $a$ ($b$) represents  Alice's (Bob's) location. After the measurements, Charlie announces his measurement results in a public channel.

Step3: Alice and Bob announce the generation bases of all their qudits in all DOFs through public channels. If they select different bases in a DOF, they have to discard the information encoding in this DOF. On the other hand, if they select the same base in a DOF, they can obtain the encoding information for each other in this DOF. Here, we suppose that the key is transmitted from Alice to Bob. To guarantee that their photon strings are correctly correlated, Bob needs to make a post-selection. The post-selection principles in three DOFs are totally the same, which are shown in Tab. 1. If both they use the rectilinear base in a DOF, Bob has to apply a bit-flip operation to his data corresponding to the measurement result of $|\psi^{\pm}\rangle$, and he keeps his original data corresponding to the measurement result of $|\phi^{\pm}\rangle$. On the other hand, if both they use the diagonal base in a DOF, Bob applies a bit-flip operation to his data corresponding to the measurement result of $|\phi^{-}\rangle$ or $|\psi^{-}\rangle$, and he keeps his original data corresponding to the measurement result of $|\phi^{+}\rangle$ or $|\psi^{+}\rangle$.

Step4: Alice and Bob repeat steps (1) to (3) until they get enough amount of keys.

Step5: To ensure the security of the key transmission, Alice and Bob choose the qudits encoding in diagonal bases $\{|+\rangle_{p}, |-\rangle_{p}\}$, $\{|+\rangle _{f}, |-\rangle_{f}\}$, and $\{|+\rangle_{s}, |-\rangle_{s}\}$ to evaluate the quantum bit error rates (QBERs) in three DOFs. In detail, they announce the
encoding information in a DOF over a public channel. This
enables them to check Charlie's measurement results by calculating
the QBER in this DOF. If the QBERs in all DOFs are smaller than the threshold, the key transmission process can be trusted. Otherwise, if the QBER in any DOF is higher than the threshold, the key transmission process has to be abandoned.

Step6: Alice and Bob perform the error correction and private  amplification  to complete the generation of final secure keys.

\begin{table}
\begin{center}
\caption{The post-selection principle of Bob in all DOFs based on the Bell-state measurement (BSM) result and the basis selection }
\begin{tabular}{|c|c|c|c|c|c|c|c|c|}
\hline
 BSM result & {$|\Phi^{+}\rangle$} & {$|\Phi^{-}\rangle $} &  {$|\Psi^{+}\rangle $} &  {$|\Psi^{-}\rangle$} \\
\hline
Rectilinear basis &  -  & -   &bit flip&bit flip\\
\hline
Diagonal basis&  -  &bit flip&-&bit flip\\
\hline
\end{tabular}
\end{center}
\end{table}

To facilitate understanding, we discuss our protocol in a specific example, where Alice's encoding information is 000 and Bob's encoding information is 101. Here, we neglect the security checking for simplicity.
According to the base choice, there are four different cases. In each case, we give an example to illustrate the protocol.
%000 100

In the first case, the bases in all DOFs of the photons sent by Alice and Bob are the same. We assume that Alice prepares {$|H\rangle_{a}$}{$|L\rangle_{a}$}{$|I\rangle_{a}$} and Bob prepares {$|V\rangle_{b}$}{$|L\rangle_{b}$}{$|E\rangle_{b}$}. Under this case, the whole photon state in Charlie's location can be written as
\begin{eqnarray}
&&|H\rangle_{a}|L\rangle_{a}|I\rangle_{a} \otimes |V\rangle_{b}|L\rangle_{b}|E\rangle_{b}\nonumber\\
&=&|H\rangle_{a}|V\rangle_{b}\otimes|L_{a}\rangle|L\rangle_{b}
\otimes |I\rangle_{a}|E\rangle_{b}\nonumber\\
&=&\frac{1}{\sqrt{2}}(|\Psi^{+}\rangle _{p}+|\Psi^{-}\rangle _{p}) \otimes
\frac{1}{\sqrt{2}}(|\Phi^{+}\rangle _{f}+|\Phi^{-}\rangle _{f}) \nonumber\\
&\otimes&
 \frac{1}{\sqrt{2}}(|\Psi^{+}\rangle _{s} +|\Psi^{-}\rangle _{s})\nonumber\\
&=&\frac{1}{2\sqrt{2}}(
   |\Psi^{+}\rangle _{p}  \otimes |\Phi^{+}\rangle _{f} \otimes |\Psi^{+}\rangle _{s}
  +|\Psi^{+}\rangle _{p}  \otimes |\Phi^{+}\rangle _{f} \otimes |\Psi^{-}\rangle _{s}\nonumber\\
  &+&|\Psi^{+}\rangle _{p}  \otimes |\Phi^{-}\rangle _{f} \otimes |\Psi^{+}\rangle _{s}
  +|\Psi^{+}\rangle _{p} \otimes |\Phi^{-}\rangle _{f} \otimes |\Psi^{-}\rangle _{s}\nonumber\\
  &+&|\Psi^{-}\rangle _{p}  \otimes |\Phi^{+}\rangle _{f} \otimes |\Psi^{+}\rangle _{s}
  +|\Psi^{-}\rangle _{p}  \otimes |\Phi^{+}\rangle _{f} \otimes |\Psi^{-}\rangle _{s}\nonumber\\
  &+&|\Psi^{-}\rangle _{p}  \otimes |\Phi^{-}\rangle _{f} \otimes |\Psi^{+}\rangle _{s}
  +|\Psi^{-}\rangle _{p}  \otimes |\Phi^{-}\rangle _{f} \otimes |\Psi^{-}\rangle _{s}).
  \end{eqnarray}
According to the measurement results and their base selection, Alice and Bob can know that their encoding in the polarization and second longitudinal-momentum DOFs are opposite and the encoding in the first longitudinal-momentum DOF is same. In this way, Bob only needs to make bit-flip operations in polarization and second longitudinal-momentum DOFs and keeps his original data in the first longitudinal-momentum DOF. Then, he can obtain the encoding keys from Alice.

In the second case, Alice and Bob choose same basis in two DOFs. For example, Alice prepares {$|H\rangle_{a}$}{$|L\rangle_{a}$}{$|I\rangle_{a}$} for 000 and Bob prepares {$|V\rangle_{b}$}{$|+\rangle_{fb}$}{$|E\rangle_{b}$} for 101. In this case, the whole photon state at Charlie's side is
 \begin{eqnarray}
&&|H\rangle_{a}|L\rangle_{a}|I\rangle_{a}\otimes|V\rangle_{b}|+\rangle_{fb}|E\rangle_{b}\nonumber\\
&=&|H\rangle_{a}|V\rangle_{b}\otimes|L\rangle_{a}|+\rangle_{fb}\otimes|I\rangle_{a}|E\rangle_{b}\nonumber\\
&=&\frac{1}{\sqrt{2}}(|\Psi^{+}\rangle_{p}+|\Psi^{-}\rangle _{p}) \otimes
  \frac{1}{2}(|\Phi^{+}\rangle _{f}+|\Phi^{-}\rangle _{f}+|\Psi^{+}\rangle _{f}\nonumber\\
  &+&|\Psi^{-}\rangle _{f})\otimes\frac{1}{\sqrt{2}}(|\Psi^{+}\rangle_{s}+|\Psi^{-}\rangle_{s})\nonumber\\
&=&\frac{1}{4}(
   |\Psi^{+}\rangle _{p}\otimes|\Phi^{+}\rangle_{f}\otimes|\Psi^{+}\rangle _{s}
  +|\Psi^{+}\rangle _{p}\otimes|\Phi^{+}\rangle_{f}\otimes|\Psi^{-}\rangle _{s}\nonumber\\
  &+&|\Psi^{+}\rangle _{p}\otimes|\Phi^{-}\rangle_{f}\otimes|\Psi^{+}\rangle _{s}
  +|\Psi^{+}\rangle _{p}\otimes|\Phi^{-}\rangle_{f}\otimes|\Psi^{-}\rangle _{s}\nonumber\\
&+&|\Psi^{+}\rangle _{p}\otimes|\Psi^{+}\rangle _{f}\otimes|\Psi^{+}\rangle _{s}
  +|\Psi^{+}\rangle _{p}\otimes|\Psi^{+}\rangle _{f}\otimes|\Psi^{-}\rangle _{s}\nonumber\\
 &+&|\Psi^{+}\rangle _{p}\otimes|\Psi^{-}\rangle _{f}\otimes|\Psi^{+}\rangle _{s}
  +|\Psi^{+}\rangle _{p}\otimes|\Psi^{-}\rangle _{f}\otimes|\Psi^{-}\rangle _{s}\nonumber\\
&+&|\Psi^{-}\rangle _{p}\otimes|\Phi^{+}\rangle _{f}\otimes|\Psi^{+}\rangle _{s}
  +|\Psi^{-}\rangle _{p}\otimes|\Phi^{+}\rangle _{f}\otimes|\Psi^{-}\rangle _{s}\nonumber\\
  &+&|\Psi^{-}\rangle _{p}\otimes|\Phi^{-}\rangle _{f}\otimes|\Psi^{+}\rangle _{s}
  +|\Psi^{-}\rangle _{p}\otimes|\Phi^{-}\rangle _{f}\otimes|\Psi^{-}\rangle _{s}\nonumber\\
&+&|\Psi^{-}\rangle _{p}\otimes|\Psi^{+}\rangle _{f}\otimes|\Psi^{+}\rangle _{s}
  +|\Psi^{-}\rangle _{p}\otimes|\Psi^{+}\rangle _{f}\otimes|\Psi^{-}\rangle _{s}\nonumber\\
  &+&|\Psi^{-}\rangle _{p}\otimes|\Psi^{-}\rangle _{f}\otimes|\Psi^{+}\rangle _{s}
  +|\Psi^{-}\rangle _{p}\otimes|\Psi^{-}\rangle _{f}\otimes|\Psi^{-}\rangle _{s}).\nonumber\\
  \end{eqnarray}
In this situation, after the HBSA, Bob can obtain that Alice's encoding in the polarization and second longitudinal-momentum DOFs are opposite with his encoding. However, Bob needs to abandon the information in the first longitudinal momentum DOF. The reason is that the BSA in the first longitudinal-momentum DOF may obtain all the four results with the same probability and Bob may obtain wrong result of Alice's encoding with the probability of 50\%. 

In the third case, Alice and Bob choose the same base in only one DOF. Assuming that Alice prepares {$|H\rangle_{a}$}{$|L\rangle_{a}$}{$|I\rangle_{a}$} representing 000 and Bob prepares {$|V\rangle_{b}$}{$|+\rangle_{fb}$}{$|-\rangle_{sb}$} for 101. The HBSA process of the whole photon state can be written as
\begin{eqnarray}
&&M=|H\rangle_{a}|L\rangle_{a}|I\rangle_{a}\otimes|V\rangle_{b}|+\rangle _{fb}|-\rangle_{sb}\nonumber\\
&=&|H\rangle_{a}|V\rangle_{b}\otimes|L\rangle_{a}|+\rangle _{fb}\otimes|I\rangle_{a}|-\rangle_{sb}\nonumber\\
&=&\frac{1}{\sqrt{2}}(|\Psi^{+}\rangle_{p}+|\Psi^{-}\rangle _{p})\otimes
  \frac{1}{2}(|\Phi^{+}\rangle _{f}+|\Phi^{-}\rangle _{f}+|\Psi^{+}\rangle _{f}\nonumber\\
  &+&|\Psi^{-}\rangle _{f})
  \otimes
  \frac{1}{2}(|\Phi^{+}\rangle _{s}+|\Phi^{-}\rangle _{s}-|\Psi^{+}\rangle _{s}-|\Psi^{-}\rangle _{s})\nonumber\\
&=&\frac{1}{4\sqrt{2}}(
    |\Psi^{+}\rangle _{p}\otimes|\Phi^{+}\rangle_{f}\otimes|\Phi^{+}\rangle _{s}
 +  |\Psi^{+}\rangle _{p}\otimes|\Phi^{+}\rangle_{f}\otimes|\Phi^{-}\rangle _{s}\nonumber\\
 &-&  |\Psi^{+}\rangle _{p}\otimes|\Phi^{+}\rangle_{f}\otimes|\Psi^{+}\rangle _{s}
 -  |\Psi^{+}\rangle _{p}\otimes|\Phi^{+}\rangle_{f}\otimes|\Psi^{-}\rangle _{s}\nonumber\\
&+& |\Psi^{+}\rangle _{p}\otimes|\Phi^{-}\rangle_{f}\otimes|\Phi^{+}\rangle _{s}
 +  |\Psi^{+}\rangle _{p}\otimes|\Phi^{-}\rangle_{f}\otimes|\Phi^{-}\rangle _{s}\nonumber\\
 &-&  |\Psi^{+}\rangle _{p}\otimes|\Phi^{-}\rangle_{f}\otimes|\Psi^{+}\rangle _{s}
 -  |\Psi^{+}\rangle _{p}\otimes|\Phi^{-}\rangle_{f}\otimes|\Psi^{-}\rangle _{s}\nonumber\\
&+& |\Psi^{+}\rangle _{p}\otimes|\Psi^{+}\rangle_{f}\otimes|\Phi^{+}\rangle _{s}
 +  |\Psi^{+}\rangle _{p}\otimes|\Psi^{+}\rangle_{f}\otimes|\Phi^{-}\rangle _{s}\nonumber\\
 &-&  |\Psi^{+}\rangle _{p}\otimes|\Psi^{+}\rangle_{f}\otimes|\Psi^{+}\rangle _{s}
 -  |\Psi^{+}\rangle _{p}\otimes|\Psi^{+}\rangle_{f}\otimes|\Psi^{-}\rangle _{s}\nonumber\\
&+& |\Psi^{+}\rangle _{p}\otimes|\Psi^{-}\rangle_{f}\otimes|\Phi^{+}\rangle _{s}
 +  |\Psi^{+}\rangle _{p}\otimes|\Psi^{-}\rangle_{f}\otimes|\Phi^{-}\rangle _{s}\nonumber\\
 &-&  |\Psi^{+}\rangle _{p}\otimes|\Psi^{-}\rangle_{f}\otimes|\Psi^{+}\rangle _{s}
-  |\Psi^{+}\rangle _{p}\otimes|\Psi^{-}\rangle_{f}\otimes|\Psi^{-}\rangle _{s}\nonumber\\
&+& |\Psi^{-}\rangle _{p}\otimes|\Phi^{+}\rangle_{f}\otimes|\Phi^{+}\rangle _{s}
 +  |\Psi^{-}\rangle _{p}\otimes|\Phi^{+}\rangle_{f}\otimes|\Phi^{-}\rangle _{s}\nonumber\\
 &-&  |\Psi^{-}\rangle _{p}\otimes|\Phi^{+}\rangle_{f}\otimes|\Psi^{+}\rangle _{s}
 -  |\Psi^{-}\rangle _{p}\otimes|\Phi^{+}\rangle_{f}\otimes|\Psi^{-}\rangle _{s}\nonumber\\
&+& |\Psi^{-}\rangle _{p}\otimes|\Phi^{-}\rangle_{f}\otimes|\Phi^{+}\rangle _{s}
 +  |\Psi^{-}\rangle _{p}\otimes|\Phi^{-}\rangle_{f}\otimes|\Phi^{-}\rangle _{s}\nonumber\\
 &-&  |\Psi^{-}\rangle _{p}\otimes|\Phi^{-}\rangle_{f}\otimes|\Psi^{+}\rangle _{s}
 -  |\Psi^{-}\rangle _{p}\otimes|\Phi^{-}\rangle_{f}\otimes|\Psi^{-}\rangle _{s}\nonumber\\
&+& |\Psi^{-}\rangle _{p}\otimes|\Psi^{+}\rangle_{f}\otimes|\Phi^{+}\rangle _{s}
 +  |\Psi^{-}\rangle _{p}\otimes|\Psi^{+}\rangle_{f}\otimes|\Phi^{-}\rangle _{s}\nonumber\\
 &-&  |\Psi^{-}\rangle _{p}\otimes|\Psi^{+}\rangle_{f}\otimes|\Psi^{+}\rangle _{s}
 -  |\Psi^{-}\rangle _{p}\otimes|\Psi^{+}\rangle_{f}\otimes|\Psi^{-}\rangle _{s}\nonumber\\
&+& |\Psi^{-}\rangle _{p}\otimes|\Psi^{-}\rangle_{f}\otimes|\Phi^{+}\rangle _{s}
 + |\Psi^{-}\rangle _{p}\otimes|\Psi^{-}\rangle_{f}\otimes|\Phi^{-}\rangle _{s}\nonumber\\
 &-&|\Psi^{-}\rangle _{p}\otimes|\Psi^{-}\rangle_{f}\otimes|\Psi^{+}\rangle _{s}
 -|\Psi^{-}\rangle _{p}\otimes|\Psi^{-}\rangle_{f}\otimes|\Psi^{-}\rangle _{s}.
 ).
\end{eqnarray}
It is obvious that Bob can only obtain Alice's encoding in the polarization DOF after the bit-flip operation.

In the last case, Alice and Bob choose different bases in all DOFs. For example, Alice prepares {$|H\rangle_{a}$}{$|L\rangle_{a}$}{$|I\rangle_{a}$} state and Bob prepares {$|-\rangle _{pb}$}{$|+\rangle _{fb}$}{$|-\rangle _{sb}$}. In this case, they have to abandon the two photons.

The possible BSA results and Bob's post-selection operations in three DOFs are described in Tab. II, Tab. III, and Tab. IV, respectively.
\begin{table*}[!htbp]
\begin{center}
\caption{Possible BSA results and Bob's post-selection operation in polarization DOF. }
\resizebox{\textwidth}{25mm}{
\begin{tabular}{|c|c|p{1.5cm}<{\centering}|p{1.5cm}<{\centering}|p{1.5cm}<{\centering}|p{1.5cm}<{\centering}|c|}
\hline
\multicolumn{2}{|c}{Bit data}& \multicolumn{4}{|c|}{BSA results and probabilities}& \multicolumn{1}{c|}{Whether or not to bit-flip } \\
\cline{1-7}
Alice&Bob&$|\phi^{+}\rangle_{p}$&$|\phi^{-}\rangle_{p}$&$|\psi^{+}\rangle_{p}$&$|\psi^{-}\rangle_{p}$&\\
\cline{1-7}
$|H\rangle$& $|H\rangle$& $\frac{1}{2}$& $\frac{1}{2}$& $0$& $0$& $-$\\
\cline{1-7}
$|H\rangle$& $|V\rangle$& $0$& $0$& $\frac{1}{2}$& $\frac{1}{2}$& bit-flip\\
\cline{1-7}
$|V\rangle$& $|H\rangle$& 0& 0& $\frac{1}{2}$& $\frac{1}{2}$& bit-flip\\
\cline{1-7}
$|V\rangle$& $|V\rangle$& $\frac{1}{2}$& $\frac{1}{2}$& 0& 0& $-$\\
\cline{1-7}
$\frac{1}{\sqrt{2}}(|H\rangle+|V\rangle)$& $\frac{1}{\sqrt{2}}(|H\rangle+|V\rangle)$& $\frac{1}{2}$& 0& $\frac{1}{2}$& 0& $-$\\
\cline{1-7}
$\frac{1}{\sqrt{2}}(|H\rangle+|V\rangle)$& $\frac{1}{\sqrt{2}}(|H\rangle-|V\rangle)$& 0& $\frac{1}{2}$& 0& $\frac{1}{2}$& bit flip\\
\cline{1-7}
$\frac{1}{\sqrt{2}}(|H\rangle-|V\rangle)$& $\frac{1}{\sqrt{2}}(|H\rangle+|V\rangle)$& 0& $\frac{1}{2}$& 0& $\frac{1}{2}$& bit flip\\
\cline{1-7}
$\frac{1}{\sqrt{2}}(|H\rangle-|V\rangle)$& $\frac{1}{\sqrt{2}}(|H\rangle-|V\rangle)$& $\frac{1}{2}$& 0& $\frac{1}{2}$& 0& $-$\\
\hline
\end{tabular}}
\end{center}
\end{table*}

%\end{document}
\begin{table*}[!htbp]
\begin{center}
\caption{Possible BSA results and Bob's post-selection operation in the first longitudinal-momentum DOF. }
\resizebox{\textwidth}{25mm}{
\begin{tabular}{|c|c|p{1.5cm}<{\centering}|p{1.5cm}<{\centering}|p{1.5cm}<{\centering}|p{1.5cm}<{\centering}|c|}
\hline
\multicolumn{2}{|c}{Bit data}& \multicolumn{4}{|c|}{BSA results and probabilities}& \multicolumn{1}{c|}{Whether or not to bit-flip } \\
\cline{1-7}
Alice&Bob&$|\phi^{+}\rangle_{f}$&$|\phi^{-}\rangle_{f}$&$|\psi^{+}\rangle_{f}$&$|\psi^{-}\rangle_{f}$&\\
\cline{1-7}
$|L\rangle$& $|L\rangle$& $\frac{1}{2}$& $\frac{1}{2}$& $0$& $0$& $-$\\
\cline{1-7}
$|L\rangle$& $|R\rangle$& $0$& $0$& $\frac{1}{2}$& $\frac{1}{2}$& bit-flip\\
\cline{1-7}
$|R\rangle$& $|L\rangle$& 0& 0& $\frac{1}{2}$& $\frac{1}{2}$& bit-flip\\
\cline{1-7}
$|R\rangle$& $|R\rangle$& $\frac{1}{2}$& $\frac{1}{2}$& 0& 0& $-$\\
\cline{1-7}
$\frac{1}{\sqrt{2}}(|L\rangle+|R\rangle)$& $\frac{1}{\sqrt{2}}(|L\rangle+|R\rangle)$& $\frac{1}{2}$& 0& $\frac{1}{2}$& 0& $-$\\
\cline{1-7}
$\frac{1}{\sqrt{2}}(|L\rangle+|R\rangle)$& $\frac{1}{\sqrt{2}}(|L\rangle-|R\rangle)$& 0& $\frac{1}{2}$& 0& $\frac{1}{2}$& bit flip\\
\cline{1-7}
$\frac{1}{\sqrt{2}}(|L\rangle-|R\rangle)$& $\frac{1}{\sqrt{2}}(|L\rangle+|R\rangle)$& 0& $\frac{1}{2}$& 0& $\frac{1}{2}$& bit flip\\
\cline{1-7}
$\frac{1}{\sqrt{2}}(|L\rangle-|R\rangle)$& $\frac{1}{\sqrt{2}}(|L\rangle-|R\rangle)$& $\frac{1}{2}$& 0& $\frac{1}{2}$& 0& $-$\\
\hline
\end{tabular}}
\end{center}
\end{table*}

\begin{table*}[!htbp]
\begin{center}
\caption{Possible BSA results and Bob's post-selection operation in the second longitudinal-momentum DOF.}
\resizebox{\textwidth}{25mm}{
\begin{tabular}{|c|c|p{1.5cm}<{\centering}|p{1.5cm}<{\centering}|p{1.5cm}<{\centering}|p{1.5cm}<{\centering}|c|}
\hline
\multicolumn{2}{|c}{Bit data}& \multicolumn{4}{|c|}{BSA results and probabilities}& \multicolumn{1}{c|}{Whether or not to bit-flip } \\
\cline{1-7}
Alice&Bob&$|\phi^{+}\rangle_{s}$&$|\phi^{-}\rangle_{s}$&$|\psi^{+}\rangle_{s}$&$|\psi^{-}\rangle_{s}$&\\
\cline{1-7}
$|I\rangle$& $|I\rangle$& $\frac{1}{2}$& $\frac{1}{2}$& $0$& $0$& $-$\\
\cline{1-7}
$|I\rangle$& $|E\rangle$& $0$& $0$& $\frac{1}{2}$& $\frac{1}{2}$& bit-flip\\
\cline{1-7}
$|E\rangle$& $|I\rangle$& 0& 0& $\frac{1}{2}$& $\frac{1}{2}$& bit-flip\\
\cline{1-7}
$|E\rangle$& $|E\rangle$& $\frac{1}{2}$& $\frac{1}{2}$& 0& 0& $-$\\
\cline{1-7}
$\frac{1}{\sqrt{2}}(|I\rangle+|E\rangle)$& $\frac{1}{\sqrt{2}}(|I\rangle+|E\rangle)$& $\frac{1}{2}$& 0& $\frac{1}{2}$& 0& $-$\\
\cline{1-7}
$\frac{1}{\sqrt{2}}(|I\rangle+|E\rangle)$& $\frac{1}{\sqrt{2}}(|I\rangle-|E\rangle)$& 0& $\frac{1}{2}$& 0& $\frac{1}{2}$& bit flip\\
\cline{1-7}
$\frac{1}{\sqrt{2}}(|I\rangle-|E\rangle)$& $\frac{1}{\sqrt{2}}(|I\rangle+|E\rangle)$& 0& $\frac{1}{2}$& 0& $\frac{1}{2}$& bit flip\\
\cline{1-7}
$\frac{1}{\sqrt{2}}(|I\rangle-|E\rangle)$& $\frac{1}{\sqrt{2}}(|I\rangle-|E\rangle)$& $\frac{1}{2}$& 0& $\frac{1}{2}$& 0& $-$\\
\hline
\end{tabular}}
\end{center}
\end{table*}

\section{The key generation rate of the MDI-QKD in 3DOFs}
In this section, we discuss the security and key generation rate of our MDI-QKD protocol adopting three DOFs of single qudits. Our MDI-QKD protocol is based on the original MDI-QKD protocol from Lo \emph{et al.} \cite{MDI-QKD}. In our MDI-QKD protocol, Alice and Bob use the single photons combined
with Pol-Ms and BSs to prepare the signal photon qudits and send them to Charlie to make the HBSA. According to the security proof of the original MDI-QKD protocol, the security proof of our protocol is also based on that of the time-reversed EPR-based QKD protocol \cite{proof}.

In theory, the encoding space of our MDI-QKD protocol includes three essentially independent subspaces. In ideal scenario, say, ideal single photon sources and quantum channels, one single photon can generate three bits of secure keys. As a result, the key generation rate of our MDI-QKD protocol is $R=3$. On the other hand, if we take imperfections such as imperfect single photon sources and noisy quantum channels, the key generation rate will be reduced. Here, we investigate the influence from the misalignment in imperfect single photon sources and channels to the quantum bit error rate (QBER) and the key generation rate of our MDI-QKD protocol. In each DOF, we use the encoding in rectilinear base  for key generation and the encoding in diagonal basis to evaluate the QBER.
 According to the original MDI-QKD protocol \cite{lo1999,lo2005,Koashi2009,MDI-QKD}, the secure key generation rate in the $ith$ DOF can be written as
 \begin{eqnarray}
R_{i}=R_{0}(1-H(e_{x_{i}})-f(e_{z_{i}})H(e_{z_{i}})),\label{Ri}
  \end{eqnarray}
 where $i=p,f,s$, and $f(e_{z_{i}})$ is an inefficiency function for the error correction in the $ith$ DOF. We suppose that Charlie locates in the middle of Alice and Bob, so that the distance ($d_{AC}$) from Alice to Charlie is half of the distance from Alice to Bob ($d$).  In this way, $R_{0}=10^{-a_{0}d/20}\cdot10^{-a_{0}d/20}=10^{-a_{0}d/10}$ ($a=0.2 dB/km$). $H(x)$ is the binary entropy with the form of
 \begin{eqnarray}
 H(x)=-xlog_{2}x-(1-x)log_{2}(1-x).
 \end{eqnarray}
 As the encoding in three DOFs are independent, the total secure key generation rate can be calculated as
\begin{eqnarray}
R&=&R_{0}[(1-H(e_{x_{p}})-f(e_{z_{p}})H(e_{z_{p}}))\nonumber\\
&+&(1-H(e_{x_{f}})-f(e_{z_{f}})H(e_{z_{f}}))\nonumber\\
&+&(1-H(e_{x_{s}})-f(e_{z_{s}})H(e_{z_{s}}))].
\end{eqnarray}
If we suppose the QBERs in three DOFs are the same, $R$ is three times of $R_{i}$.

We first consider the misalignment influence from the imperfect single photon sources. We take the source misalignment in polarization DOF as an example. When Alice (Bob) aims to prepare a photon in $|H\rangle$ or $|V\rangle$, the practical photon state expands to $\beta_{p}|H\rangle+\sqrt{1-\beta^{2}_{p}}|V\rangle$ or $\beta_{p}|V\rangle+\sqrt{1-\beta^{2}_{p}}|H\rangle$, respectively, while when they aim to prepare a photon in $|\pm_{p}\rangle$, the practical photon state expands to $\beta_{p}|+_{p}\rangle+\sqrt{1-\beta^{2}_{p}}|-_{p}\rangle$ or $\beta_{p}|-\rangle+\sqrt{1-\beta^{2}_{p}}|+\rangle$, respectively \cite{highdimension10}.

Next, we analyze the most general errors affecting qudits in transmission channels, say, the polarization and spatial-mode misalignments caused by the channel noise. We define that the operator $U_{A_{p(f,s)}}$  and $U_{B_{p(f,s)}}$ represent
the misalignment in the polarization (first and second longitudinal momentum) DOF of Alice's and Bob's channel transmission. Here, we consider a simplified model with a two-dimensional unitary matrix as \cite{kok2007linear}
\begin{equation}
U_{i}=
\left(
\begin{array}{cc}
cos\theta_{i}  & -sin\theta_{i} \\
sin\theta_{i} &cos\theta_{i}
\end{array}
\right),
\end{equation}
where $i=p,f,s$, and $\theta_{i}$ is in the scale of $[-\pi, \pi]$.
We also take the polarization DOF as an example. Due to the misalignment in channel transmission, we can obtain $|H\rangle\rightarrow cos\theta_{p}|H\rangle-sin\theta_{p}|V\rangle$, $|V\rangle\rightarrow cos\theta_{p}|V\rangle+sin\theta_{p}|H\rangle$, $|+_{p}\rangle\rightarrow cos\theta_{p}|+_{p}\rangle+sin\theta_{p}|-_{p}\rangle$, $|-_{p}\rangle\rightarrow cos\theta_{p}|-_{p}\rangle-sin\theta_{p}|+_{p}\rangle$. As a result, the misalignment error rate is defined as $sin^{2}\theta_{p}$. In the other two DOFs, we can obtain similar misalignment error rates as $sin^{2}\theta_{f}$ and $sin^{2}\theta_{s}$, respectively.

 In this way, we combine the misalignment from source and channel transmission. If Alice and Bob aim to send $|H\rangle|H\rangle$ to Charlie, the practical photon states Charlie receiving will evolve to
 \begin{eqnarray}
|HH\rangle&\rightarrow&\beta_{p}^{2}|HH\rangle+(1-\beta_{p}^{2})|VV\rangle\nonumber\\
&+&\beta_{p}\sqrt{1-\beta_{p}^{2}}(|HV\rangle+|VH\rangle)\nonumber\\
&\rightarrow&A_{0p}|HH\rangle+A_{1p}|VV\rangle\nonumber\\
&+&A_{2p}(|HV\rangle+|VH\rangle)\nonumber\\
&=&\frac{A_{0p}+A_{1p}}{\sqrt{2}}|\phi^{+}\rangle+\frac{A_{0p}-A_{1p}}{\sqrt{2}}|\phi^{-}\rangle\nonumber\\
&+&\sqrt{2}A_{2p}|\psi^{+}\rangle,
\end{eqnarray}
where
\begin{eqnarray}
A_{0p}&=&\beta_{p}^{2}cos^{2}\theta_{p}+(1-\beta_{p}^{2})sin^{2}\theta_{p}\nonumber\\
&+&2\beta_{p}\sqrt{1-\beta_{p}^{2}}cos\theta_{p}sin\theta_{p}, \nonumber\\
A_{1p}&=&\beta_{p}^{2}sin^{2}\theta_{p}+(1-\beta_{p}^{2})cos^{2}\theta_{p}\nonumber\\
&-&2\beta_{p}\sqrt{1-\beta_{p}^{2}}cos\theta_{p}sin\theta_{p}, \nonumber\\
A_{2p}&=&-\beta_{p}^{2}cos\theta_{p}sin\theta_{p}+(1-\beta_{p}^{2})cos\theta_{p}sin\theta_{p}\nonumber\\
&+&\beta_{p}\sqrt{1-\beta_{p}^{2}}(cos^{2}\theta_{p}-sin^{2}\theta_{p}).
\end{eqnarray}
In this way, there is a probability of $2A_{2p}^{2}$ that Charlie obtains wrong BSA result $|\psi^{+}\rangle$, thus increasing the QBER $e_{z_{p}}=2A_{2p}^{2}$. If Alice and Bob both select the diagonal basis, i.e., $|+\rangle_{p}|+\rangle_{p}$, they will also obtain error rate $e_{xp}=e_{zp}$.  Similarly, we can also calculate the QBER in other two DOFs  as $e_{z_{f}}=e_{x_{f}}=2A_{2f}^{2}$ and $e_{z_{s}}=e_{x_{s}}=2A_{2s}^{2}$, respectively.

\begin{figure*}%[tpb]
\begin{center}
\includegraphics[width=12cm,angle=0]{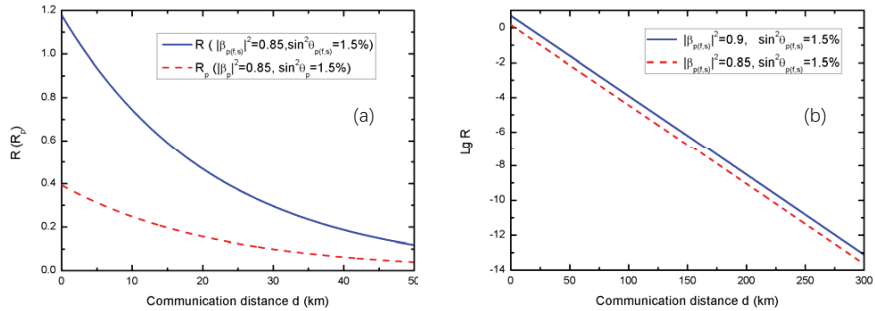}
\caption{(a) The key generation rate ($R_{p}$) in polarization DOF and total secure key generation rates ($R$) of our  MDI-QKD protocol as a function of the communication distance between Alice and Bob.  We control $|\beta_{p(f,s)}|^{2}=0.85$ and $sin^{2}\theta_{p(f,s)}=1.5\%$. The inefficiency function for the error correction process is set as $f_{e_{z_{p(f,s)}}}=1$ \cite{highdimension10}. (b) The value of $Lg R$ as a function $d$ in the scale of $0\leq d\leq300 km$, with $sin^{2}\theta_{p(f,s)}=1.5\%$ and $|\beta_{p(f,s)}|^{2}=0.85$ and $0.9$.}
\end{center}
\end{figure*}

Taking the value of QBER in each DOF into Eq. (\ref{Ri}), we can calculate the key generation rate ($R_{p(f,s)}$) in each DOF and the total key generation rate ($R$). Here, we control  $|\beta_{p(f,s)}|^{2}=0.85$ and $sin^{2}\theta_{p(f,s)}=1.5\%$. The inefficiency function for the error correction process is set to $f_{e_{z_{p(f,s)}}}=1$. In Fig. 2(a), we show the values of  $R$ and $R_{p}$ as a function of the communication distance (d) between Alice and Bob in the scale of $0\leq d\leq50 km$. Actually, $R_{p}$ equals to the key generation rate of the original MDI-QKD using only one DOF of single photons. It can be found that by encoding in three DOFs of single photons, we can efficiently increase the key generation rate of MDI-QKD. In Fig. 2(b), we show the value of $Lg R$ as a function $d$ in the scale of $0\leq d\leq300$ $km$. Here, we control $sin^{2}\theta_{p(f,s)}=1.5\%$ and $|\beta_{p(f,s)}|^{2}=0.85$ and $0.9$. It is obvious that with the growth of QBERs, the total key generation rate obviously reduces.

\section{Discussion and Conclusion}
In the paper, we propose a high-capacity MDI-QKD protocol, which adopts the photon qudits encoding in three DOFs to generate keys. In our protocol, both Alice and Bob generate single photons encoding in polarization and two longitudinal-momentum DOFs and send them to Charlie for HBSA. Here, we adopt the HBSA protocol in Ref. \cite{HBSA} to completely distinguish the 64 hyperentangled Bell states with the help of the QND gates constructed by cross-Kerr nonlinearity. As the encoding information in each DOF can be used to generate secure key, encoding in multiple DOFs of single photons can efficiently increase the key generation rate of MDI-QKD. Actually, our MDI-QKD protocol can be extended to use general $N$ DOFs of single photons ($N\geq2$). In this general case, both Alice and Bob generate single photons encoding in $N$ DOFs. In each DOF, they randomly choose rectilinear or diagonal base to encode the photon. The four Bell states in each DOF can be written as
\begin{eqnarray}
|\Phi^{\pm}\rangle _{k}&=&\frac{1}{\sqrt{2}}(|0\rangle_{a}|0\rangle_{b}\pm|1\rangle_{a}|1\rangle_{b}),\nonumber\\
 |\Psi^{\pm}\rangle _{k}&=&\frac{1}{\sqrt{2}}(|0\rangle_{a}|1\rangle_{b}\pm|1\rangle_{a}|0\rangle_{b}).
\end{eqnarray}
where $|0\rangle$ and $|1\rangle$ represent two orthogonal states in rectilinear or diagonal base in the $kth$ DOF ($k=1,2,3,\cdots N$). In the HBSA process, Charlie requires to construct the QND gates, which make he can completely distinguish all the four Bell states in each DOF, that is, $4^{N}$ hyperentangled Bell states in total.
As long as the QND in each DOF is available, we can calculate the key generation rate in each DOF according to Eq. (\ref{Ri}).
Therefore, the total key generation rate of our MDI-QKD protocol using $N$ DOFs of single photons can be expressed as
\begin{eqnarray}
R&=&NR_{p}.
\end{eqnarray}

Here, we compare our protocol with the $N$-dimensional MDI-QKD protocol put forward by Dellantonio \emph{et al.} \cite{highdimension10}. In Ref. \cite{highdimension10}, the MDI-QKD protocol uses the HD qudits encoding in different spatial paths or time slots, which could also increase MDI-QKD's key generation rate. Our protocol has an attractive advantage over theirs, that is, our protocol is more flexible. In detail, for the protocol in Ref. \cite{highdimension10}, if a detection failure or an error occurs in a qudit, this qudit cannot be used to generate the key. For our MDI-QKD protocol, the HD system includes $N$ essentially independent subsystems. The $N$ subsystems can be manipulated and measured independently. As a result, if a detection failure or an error occurs in one DOF, the encoding information in this DOF has to be discarded. However, the encoding information in other DOFs can also be used to generate the key. We have to abandon a single photon qudit only when the error or detection efficiency happens in all DOFs. In this way, the more DOFs of single photons the MDI-QKD protocol uses, the MDI-QKD protocol shows the stronger resistance to errors and detection inefficiency. On the other hand, our MDI-QKD protocol requires single-photon sources. In practical experiments, the phase-randomized coherent state source is often used instead of the traditional single-photon source. However, the secret key rate will decrease significantly due to the vacuum state with large probability. For increasing the secret key rate, we may also adopt the coherent-state superpositions source, which can avoid the disadvantage of the phase-randomized coherent state source and has been used in MDI-QKD field \cite{MDI-QKD10}.

The key step of our multiple-DOF MDI-QKD protocol is the HBSA. Such HBSA protocol was also used in realizing the high-capacity quantum secure direct communication \cite{hyperQSDC} and constructing the two-photon three DOFs hyper-parallel controlled phase-flip gate \cite{hypergate}. However, the complete HBSA protocol adopted in our MDI-QKD protocol requires the cross-Kerr nonlinearity technique, which is still a challenge under current experimental condition. The main reason is that the natural cross-Kerr nonlinearity is quite weak, so that it is difficult to use homodyne detection to discriminate two overlapping coherent states \cite{kok2007linear}. However, some great progresses on the cross-Kerr nonlinearity have been made in recent years \cite{cross2,cross3,cross5,cross6,kerr}. For example, in 2016, D\"{u}rr \emph{et al.} reported that they realized a strong interaction in Rydberg electromagnetically induced transparency (EIT) experiments to create a large controlled phase shift of $3.3\pm 0.2$ rad, with the incoming control pulses containing an average of 0.6 photons \cite{cross6}. In 2019, the group of Sinclair observed the nonlinear phases of 8 $mrad/nW$ of signal power, corresponding to a $\chi^{3}$ of $10^{-8}$ $m^{2}/V^{2}$ \cite{kerr}. In addition,  several attractive HBSA protocols based on other nonlinear optical elements have also been proposed, such as quantum-dot spins in optical microcavities \cite{ren2012complete,wang2012generation,wang2016error}, nitrogen-vacancy centers in resonators \cite{liu2015generation}, and a three-level $\Lambda$-type system \cite{wang2016complete}. Recently, based on probabilistic HBSA, the first quantum teleportation of a single photon encoded in both spin and orbital angular momentum DOFs has also been achieved by the group of Pan \cite{hypercommun2}. Therefore, based on these works, we believe that the $N$-DOF complete HBSA may be experimentally realized in the future.

\section{Conclusion}
In summary, we propose a high-capacity MDI-QKD protocol, which adopts single photons encoding in multiple DOFs to generate secure keys. We take the qudits encoding in polarization and two longitudinal momentum DOFs as an example. Alice and Bob randomly prepare the qudits encoding in above three DOFs and send them to a third party Charlie through quantum channels. Then, Charlie makes the complete HBSA on the photon pairs and publishes his results. According to Charlie's HBSA results and the bases of photon qudits in three DOFs announced by both sides, Bob can obtain Alice's secure key when they choose same base. Our MDI-QKD protocol is unconditionally secure. By adopting the qudits encoding in multiple DOFs, our protocol can obviously increase the information capacity of single photon, and thus increase MDI-QKD's key generation rate. Moreover, comparing with previous HD MDI-QKD protocol in one DOF, our protocol still has some advantages. The HD system in our protocol includes some essentially independent subsystems, which can be manipulated and measured independently. As a result, a detection failure or an error  in one DOF does not affect the encoding information in other DOFs, which can also be used to generate the key. In this way, our MDI-QKD protocol is more flexible. The more DOFs we adopt, our MDI-QKD protocol has the stronger resistance to errors or detection inefficiency. Based on above features, our MDI-QKD protocol may have application potential in the near future.

\acknowledgements
This work was supported by the National Natural Science  Foundation of China under Grant  No. 11974189, the Postgraduate Research $\&$  Practice
Innovation Program of Jiangsu Province under Grant No. SJCX19-0241, and a Project Funded by the Priority
Academic Program Development of Jiangsu Higher Education Institutions.

\end{document}